\documentclass[aps,twocolumn]{revtex4}
\usepackage{amssymb}
\usepackage{amsmath}
\usepackage{epsfig}
\begin{document}

\title{Gap soliton formation by nonlinear supratransmission in Bragg media}
\author{ J. Leon}
\affiliation{Physique Math\'ematique et Th\'eorique, CNRS-UMR5825\\
Universit\'e Montpellier 2, 34095 MONTPELLIER  (France)}
\author{ A. Spire}
\affiliation{Centro de Fisica da Materia Condensada,\\
 Universidade de Lisboa, 1649-003 LISBOA (Portugal)}

\begin{abstract}A Bragg medium in the nonlinear Kerr regime, submitted to
incident cw-radiation at a frequency in a band gap, switches from total
reflection to  transmission when  the incident energy overcomes some threshold.
We demonstrate that this is a  result of  {\em nonlinear supratransmission}, 
which allows to prove that : i) the threshold incident amplitude is simply
expressed in terms of the deviation from the Bragg resonance, ii) the process
is not the result of a shift of the gap in the nonlinear dispersion relation,
iii) the transmission does occur by means of gap soliton trains, as
experimentally observed [D. Taverner {\em et al.}, Opt Lett 23 (1998) 328], iv)
the required energy tends to zero close to the band edge.\end{abstract}
\maketitle

\paragraph*{Introduction.}

Light propagation in dielectric media with periodically varying index acquires
intriguing properties when Kerr nonlinearities come into play.  From a linear
theory \cite{Brill} we know that such a medium is a {\em photonic band gap}
structure, or Bragg medium, which is totally reflective when the frequency of
the incoming wave falls in one of the gaps.  

However, for sufficiently energetic irradiation, the Kerr effect becomes
sensible and the Bragg medium may become transparent in the gap, switching from
total reflection to  {\em high transmissivity}, which has been predicted in
1979 \cite{winful} by using the stationary coupled mode approach of \cite{kog},
and which has been experimentally realized for the first time in 1992
\cite{sankey}.  

In 1987 the existence of effective wave propagation in a gap has been
associated to the existence of {\em gap solitons} derived from slowly varying
envelope limits of the Maxwell equation in a periodic structure
\cite{chen,mills}.  For unidirectionnal propagation, fast Kerr nonlinearity and
slowly varying evelope approximation, one of these limit models is the {\em
coupled mode system} \cite{desterke} which describes the evolution of the
envelopes of the contrapropagating electric fields. A rigorous derivation of
from the anharmonic Maxwell-Lorentz equations by the method of multiple scales
is given in \cite{goodman}. The explicit soliton-like solution of the coupled
mode equation has been discovered in 1989 \cite{christo}, and at the same time
generalized to a two-parameter family in \cite{aceves}.  This two-parameter gap
soliton solution, that we use here, has motivated many experimental searches of
the {\em Bragg soliton}.

After the pioneering experiments of \cite{sankey} in steady-state regime, the
first experimental observations of nonlinear pulse {\em propagation} in a fiber
Bragg grating have been performed in 1996 \cite{eggleton}  under  laser pulse
irradiation at a frequency near (but not inside) the photonic band gap.  In
that case the nonlinearity acts as a source of {\em modulationnal instability}
generating trains of pulses \cite{eggleton2}.  Particularly interesting is the
observation of pulse propagation at a velocity less than the light velocity. 
Later in 1998 \cite{taverner} a convincing experiment with a fiber Bragg
grating demonstrated the repeated formation of gap soliton under quasi-constant
wave irradiation {\em inside} a band gap, this is the problem we are interested
in.  

There has been a great deal of discussions concerning the exact role of the gap
soliton in the switching to high transmissivity, see e.g. \cite{souk}.  A
rather natural intuition  is that the nonlinearity  shifts the gap and hence
allows the medium to become transparent and to create a {\em gap soliton}. We
shall discover that this intuition is wrong as the process at the origin of
high transmissivity in Bragg media is the {\em nonlinear supratransmission}
recently discovered in the sine-Gordon chain \cite{jg-alex,nst-prl} and
experimentally realized (together with application to Josephson junctions
arrays) in \cite{nst-jpc}. As a consequence, the switching to high
transmissivity is indeed accomplished by means of gap soliton generation as a
result of a fundamental instability \cite{instab} which are effectively the
objects experimentally detected in the output in \cite{taverner}.

This result allows us to predict analytically by formula \eqref{threshold} the
threshold of incident energy of a cw-radiation above which high transmissivity
is reached, as an explicit simple function of the departure (denoted by the
dimensionless angular frequency $\Omega$) from the Bragg frequency (gap
center). In particular we obain that the process requires less energy for
frequencies close to the band edge $\Omega=1$, a result which was previously
attributed to pulse shape properties.  Interestingly enough, in the limit
$\Omega\to 1$ the required energy vanishes, opening the way to experimental
realizations.

\paragraph*{The model and its solution.}

Our starting point is the basic coupled mode system \cite{desterke,goodman}
governing the forward $E_f$ and backward $E_b$ slowly varying  envelopes of 
the electric field 
\begin{equation}
E(Z,T)=[E_fe^{ik_0Z}+E_be^{-ik_0Z}]e^{-i\omega_0 T}\ .
\end{equation}
$E(Z,T)$ is the transverse polarized component propagating in the
direction $Z$ at frequency close to the Bragg frequency $\omega_0$.  Following
\cite{goodman} we  write the coupled mode  system in reduced units as
\begin{eqnarray}\label{coup-sys}
i[\frac{\partial e}{\partial t}+\frac{\partial e}{\partial z}]+ f 
+(\frac12|e|^2+|f|^2)e=0\ ,\nonumber\\
i[\frac{\partial f}{\partial t}-\frac{\partial f}{\partial z}]+ e 
+(\frac12|f|^2+|e|^2)f=0\ .
\end{eqnarray}
The reduced units are $z=\kappa Z$ and $t=\kappa c T$ where $c$ is the light
velocity in the medium and $\kappa$ the coupling constant. The reduced field
variables are $e=E_f\sqrt{2\Gamma/\kappa}$ and $f=E_b\sqrt{2\Gamma/\kappa}$
where $\Gamma$ is the nonlinear factor. The (linear) dispersion relation of 
\eqref{coup-sys}, namely $\omega^2=1+k^2$, possess the gap $[-1,+1]$.  

The solitary wave solution of \eqref{coup-sys} given in \cite{aceves} 
has a simple stationnary expression
\begin{align}
e=\sqrt{2/3}
\lambda \,e^{-i(\Omega t -\phi)}\cosh^{-1}[\lambda(z-z_0)-\frac i2q]
\ ,\label{e-sol}\\
f=-\sqrt{2/3}
\lambda \,e^{-i(\Omega t -\phi)}\cosh^{-1}[\lambda(z-z_0) +\frac i2q]
\ .\label{f-sol}\end{align}
The real valued parameters determining this {\em `gap soliton'} are $\Omega$
(frequency), $\phi$ (initial phase) and $z_0$ (center), and we have
\begin{equation}\Omega^2=1-\lambda^2\ ,\quad \Omega=\cos q\ .\end{equation}
Then $\lambda$ plays the role of the {\em `wave number'} of the
evanescent wave.

\paragraph*{Nonlinear supratransmission threshold.}

For a Bragg medium, extending in the region $z\in[0,L]$, and initially 
{\em in the dark}, we set the initial data
\begin{equation}\label{init}
e(z,0)=0\ ,\quad f(z,0)=0\ .
\end{equation}
The boundary value problem  that mimics the scattering of an incident 
radiation at frequency $\omega_0+\kappa c\Omega$ on the medium in $z=0$ is 
\begin{equation}\label{bound}
e(0,t)=A\,e^{-i\Omega t}\ ,\quad f(L,t)=0\ ,
\end{equation}
where the second requirement means no backward wave incident from the right
in $z=L$.  The constant $A$ (in general complex valued) is the dimensionless
amplitude of the incoming radiation.  Equations \eqref{init} and \eqref{bound}
constitute a well posed initial-boundary value problem for the PDE
\eqref{coup-sys}.

In the linear case such a boundary forcing would simply generate the evanescent
wave $e(z,t)=A\,e^{-i\Omega t-\lambda z}$. In the nonlinear case however this
boundary forcing generates the solution \eqref{e-sol} for the value of $\phi$
and $z_0<0$ such that the amplitude in $z=0$ be precisely $A$. Then,
for each fixed forcing frequency $\Omega$, there exists a maximum value $A_s$
of $A$ beyond which there is no solution $\{\phi,z_0\}$. This threshold
$A_s$ is given by $|e(0,t)|$ from \eqref{e-sol} evaluated at $z_0=0$, namely by
\begin{equation}\label{threshold}
A_s=2\sqrt{2/3}\,\sin(\frac12\arccos \Omega)\ .\end{equation}
This is the threshold amplitude of the incident envelope above which the system
develops an instability and generates a propagating nonlinear mode.

\begin{figure}[ht]
\centerline{\epsfig{file=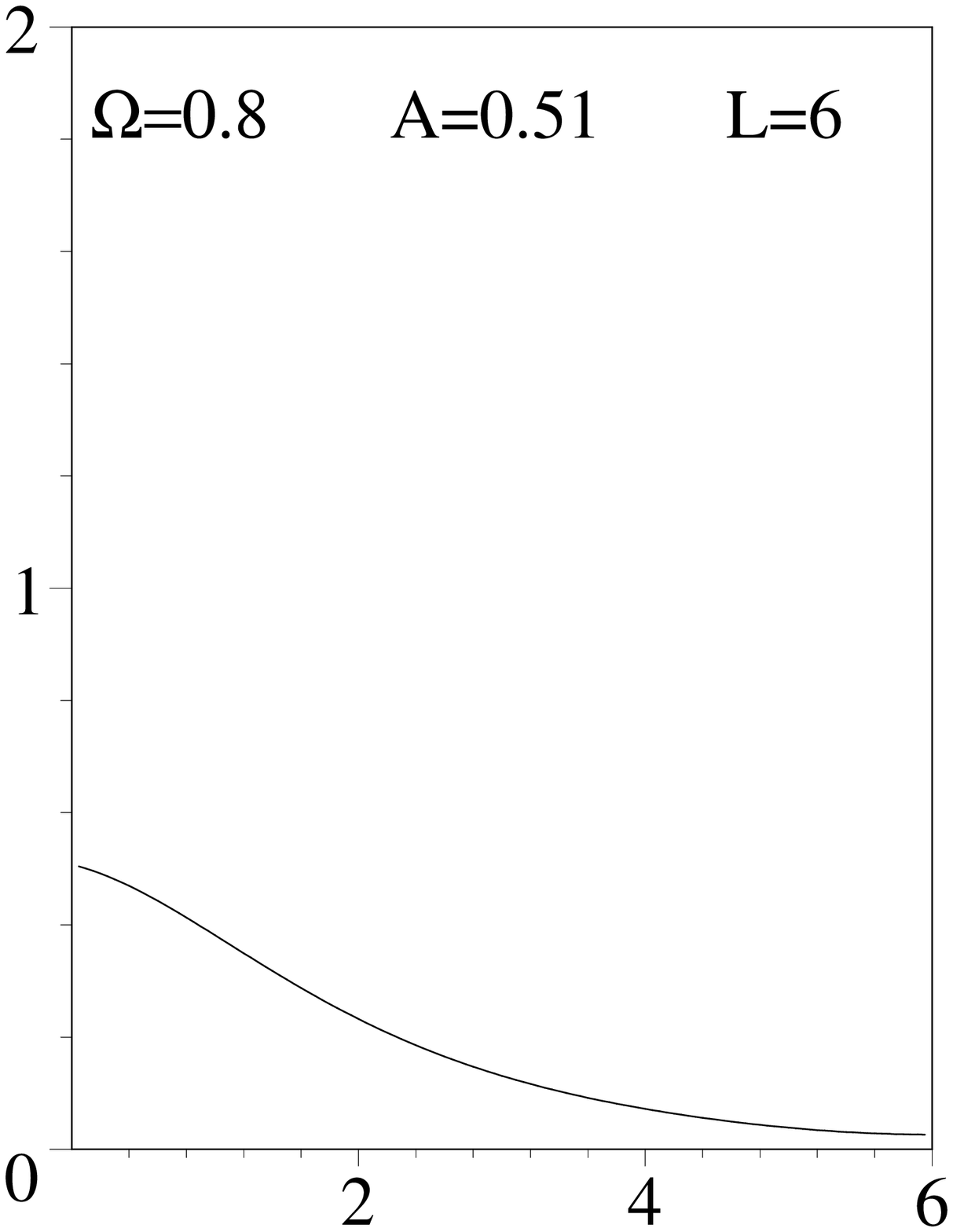,height=3cm,width=3cm}
\epsfig{file=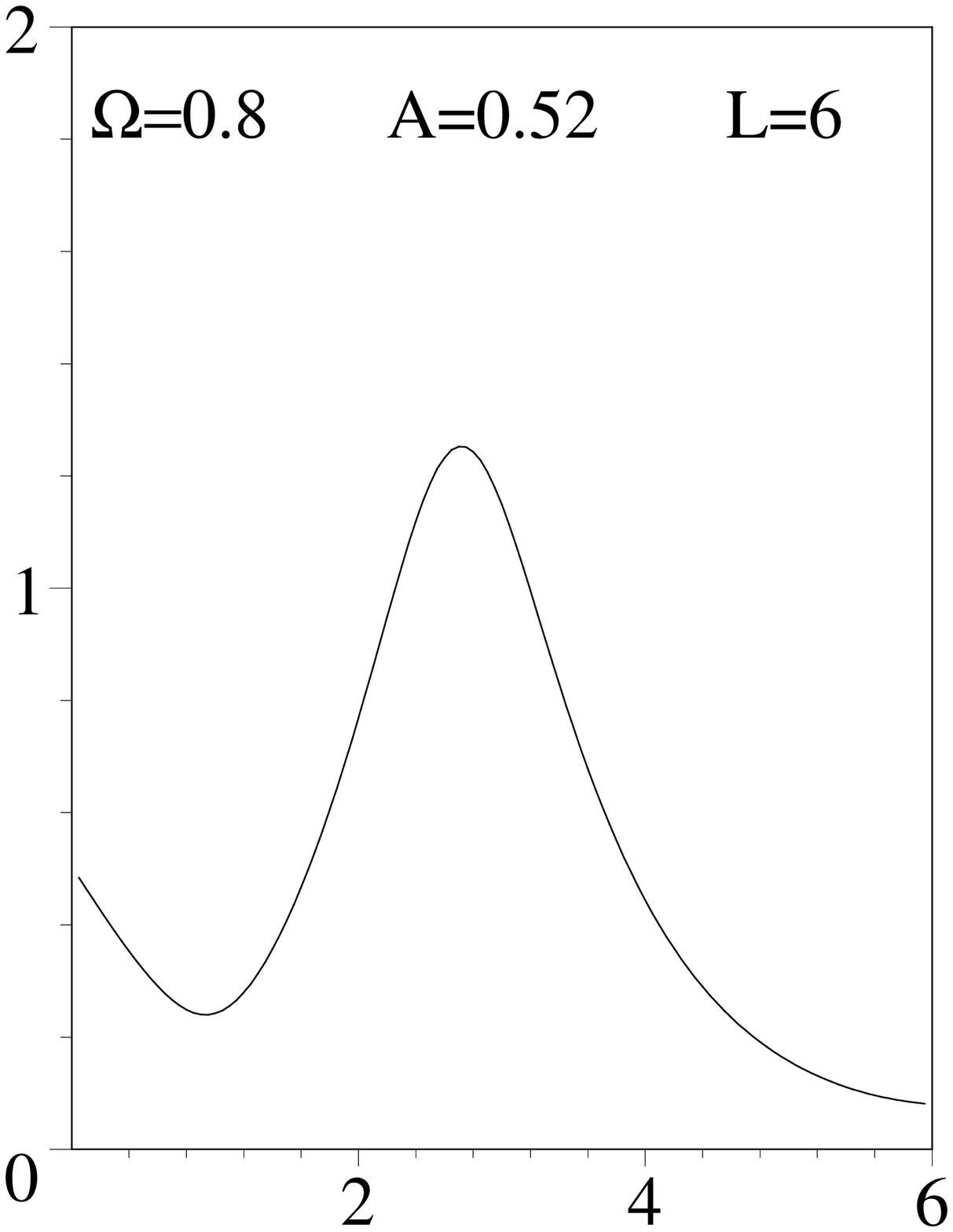,height=3cm,width=3cm}}
\caption{Plots of the amplitude $|e(z,t)|$ of the right-going envelope at 
given time ($t=145$) for two values of the driving amplitude $A$ at frequency
$\Omega=0.8$.}\label{fig:breather}\end{figure}

\paragraph*{Nonlinear dispersion relation prediction.}

 By seeking a solution of \eqref{coup-sys} $e=a\exp[i(kz-\omega t)]$,
$f=b\exp[i(kz-\omega t)]$ one gets a nonlinear algebraic system for the
unknowns $a$ (incident amplitude) and $\omega$ (incident frequency) expressed
in terms of the wave number $k$ once the amplitude $b$ has been eliminated. 
This gives the {\em nonlinear dispersion relation} $\omega(a,k)$ as a solution
of a third order algebraic equation. One of the 3 solutions must be discarded
as it is singular in the linear limit $a\to0$, the other two giving the
deformations of the linear branches $\omega(k)=\pm\sqrt{1+k^2}$.  It can be
shown that they correspond to the relations $b=-a$ (upper branch) and $b=a$
(lower branch).

Being interested in the shift of the gap, it is sufficient to study the
solutions at $k=0$ where the system simplifies, and to stick with the upper
branch (as we look at right-going incident waves) for which $b=-a$. Then
\eqref{coup-sys} readily provides the value $\omega(0,a)=1-\frac32\,a^2$ 
of the gap opening. 

For our purpose it is more convenient to express the
value $A_m$ of the incident amplitude $a$ for which the incident frequency
$\Omega$ touches the nonlinear gap edge $\omega(0,a)$, namely
\begin{equation}\label{nln-predic}
A_m=\sqrt{\frac23(1-\Omega)}\ .\end{equation}
This formula provides a prediction of {\em transparency} by nonlinear shift
of the gap. It is now compared to our prediction \eqref{threshold} by means
of numerical simulations.

\paragraph*{Numerical simulations.}

The system \eqref{coup-sys} is solved by means of an semi-implicit third order
finite difference scheme and boundary values are taken into account at first
order. Explicitely we first rewrite  \eqref{coup-sys} for two functions
$u(z,t)$ and $v(z,t)$ and replace the operators $\partial_t +\partial_z$ by the
set of differences for $u_n(t)=u(nh,t)$
\begin{align*}
& \dot u_1 + \frac1h(u_1-u_0),\quad
\dot u_2+\frac1{2h}(u_2-u_0),\quad\cdots \\
& \dot u_n+\frac1{h}\left[\frac23(u_{n+1}-u_{n-1})-
\frac1{12}(u_{n+2}-u_{n-2})\right],\quad\cdots\\
& \dot u_{N-1}+\frac1{2h}(u_{N+1}-u_{N-1}),\quad
 \dot u_{N}+\frac1{h}(u_{N}-u_{N-1}),\end{align*}
so as for $v_n(t)=v(nh,t)$. The length is $L=hN$ and  overdot means time
differentiation. 
Equation \eqref{coup-sys} results as a system of $2 N$ coupled ODE then solved
through the subroutine {\tt dsolve} of the {\tt MAPLE8} software package that
uses a Fehlberg fourth/fifth order Runge-Kutta method.  Finally the solution,
e.g. $e(z,t)$, is obtained from $u_n(t)$ by $e(z,t)=[u_{n+1}(t)+u_n(t)]/2$. 
Although such code sends some numerical noise in the solution, it is
uneffective for sufficiently small $h$ depending on the required time of
integration. 

Next, in order to avoid initial shock, the boundary condition $e(0,t)$ is
smoothly turned on and smoothly turned off by assuming instead of \eqref{bound}
\begin{equation}\label{bound-smooth}
e(0,t)=\frac A2\left[\tanh(p(t-t_0))-\tanh(p(t-t_1))\right]\ .\end{equation}
A practical interest of such an incident wave is that it reproduces the
{\em quasi-constant wave} irradiation of the experiments of \cite{taverner}.

Most of the results presented here are obtained with $N=120$ spatial mesh
points, $h=0.05$ grid spacing over (hence a length $L=6$ normalize units) a
time of integration $t_m=200$. The parameters of the boundary field are a
real-valued amplitude $A$, a slope $p=0.2$, an ignition time  $t_0=20$ and an
extiction time $t_1=180$.

We display in figure \ref{fig:breather} an instance of two different numerical
solutions (the modulus of the right-going envelope) for an incident frequency
$\Omega=0.8$ and amplitudes $A=0.51$ (no transmission) and $A=0.52$ (gap
soliton generation) when the theoretical threshold predicted by
\eqref{threshold} is $A_s=0.5164$.

\begin{figure}[ht]
\centerline{\epsfig{file=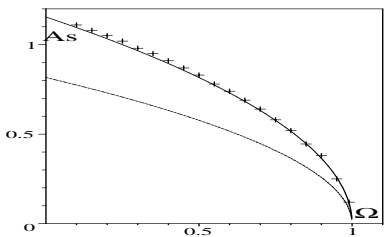,height=3cm,width=5cm}}
\caption{Values of the threshold of nonlinear supratransmission
observed on numerical simulations (crosses)
as compared with expression \eqref{threshold} (solid line) and
with \eqref{nln-predic} (dashed line).}
\label{fig:bif}\end{figure}

Using this simple diagnostic, the bifurcation predicted by expression
\eqref{threshold} is numerically checked for a series of frequency values (in
the range $[0.1,0.995]$) and we obtain the figure \ref{fig:bif} where the dots
represent the smallest value (with absolute precision of $10^{-2}$) of the
amplitude $A$ for which nonlinear supratransmission is seen to occur.  
The expression \eqref{nln-predic} is also plotted (dashed line) which shows
that the nonlinear shift provides a wrong answer.

\paragraph*{Bifurcation of transmitted energy.}

The system \eqref{coup-sys} possess the conservation law
\begin{equation}\label{cons}
\partial_t(|e|^2+|f|^2)+\partial_z(|e|^2-|f|^2)=0.\end{equation} 
 For a given boundary  condition \eqref{bound} (i.e. for fixed $A$ and
$\Omega$), we define the incident ($I$), reflected ($R$) and transmitted ($T$)
energies at given arbitrary time $t_m$ by 
\begin{align} I(A,\Omega)=\int_0^{t_m}
dt\,|e(0,t)|^2,\\ R(A,\Omega)=\int_0^{t_m} dt\,|f(0,t)|^2,\\
T(A,\Omega)=\int_0^{t_m} dt\,|e(L,t)|^2.\end{align} 
Then, upon integration of \eqref{cons} on the length $[0,L]$ and time $[0,t_m]$
we obtain
\begin{equation} R+T-I+\int_0^Ldz\,\left(|e(z,t_m)|^2+|f(z,t_m)|^2\right)=0\ . 
\end{equation}
If $t_m$ is larger than the irriadiation duration, the energy injected
eventually radiates out completely, namely $e(z,t_m)$ and $f(z,t_m)$ vanish,
and we are left with the conservation relation $R+T=I$ which can be written 
\begin{equation} \rho(A,\Omega)+\tau(A,\Omega) =1,\quad \rho=R/I,\quad
\tau=T/I.  \end{equation}

\begin{figure}[ht]
\centerline{\epsfig{file=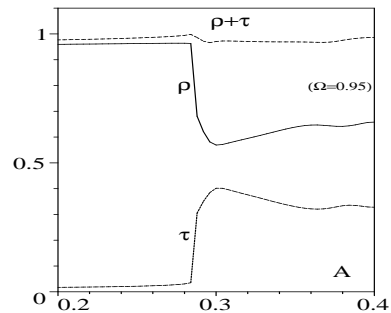,height=4cm,width=5cm}}
\caption{Plot of the reflection factor $\rho$ (full line), transmission $\tau$
(dashed line) and sum $\rho+\tau$ (dotted line) for system \eqref{coup-sys}
with boundary value \eqref{bound-smooth} and parameter values given there.}
\label{fig:energy}\end{figure}

The figure \ref{fig:energy} shows a typical numerical simulation where, for
$\Omega=0.95$, we have computed the reflection and transmission factors,
together with their sum, for 50 different values of the amplitude $A$ in
the range $[0.20,\, 0.45]$.
These simulations  showt the sudden energy flow through the medium, as a result
of nonlinear supratransmission.  Note from the sum $\rho+\tau$ before the
bifurcation on figure \ref{fig:energy} that the numerical code conserves the 
energy with a precision of $2\%$.

\begin{figure}[ht]
\centerline{\epsfig{file=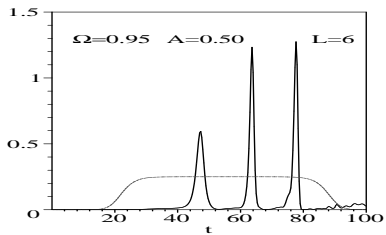,height=3cm,width=5cm}}
\caption{Energy density output $|e(L,t)|^2$ (full line) compared to the
input (dashed line) for $\omega=0.95$ and $A=0.5$.}
\label{fig:outflux}\end{figure}

The transmissivity is due to the generation and propagation of light pulses
shown figure \ref{fig:outflux} representing the
energy density $|e(L,t)|^2$ measured at the output as a function of time.
We have also plotted the input energy density given in \eqref{bound-smooth}
with parameters $t_m=100$, $t_0=20$, $t_1=90$ and $p=2$.
We show now that these are {\em gap solitons} travelling at a fraction
of the light velocity (slow light pulses).

\paragraph*{Travelling gap solitons.}

Although the system \eqref{coup-sys} is not Lorentz invariant, the propagating
solution of \cite{aceves} can still be written in terms of the boosted
variables \cite{barash}. As we are intersted  here only in the energy flux, we
write the soliton solution $|e(z,t)|^2$ moving at velocity $v<1$ with frequency
$\cos q$ as 
\begin{align}\label{e-mov}
|e(z,t)|^2=\frac{2\sin^2 q\,\left(\frac{1-v^2}{3-v^2}\right)
\left(\frac{1+v}{1-v}\right)^{1/2}}
{\left|\cosh\left(\frac{\sin q}{\sqrt{1-v^2}}(z-z_0-vt)-\frac i2q\right)
\right|^{2}}.
\end{align}
Such an expression allows to fit a given simulation by seeking the two
parameters $v$ and $q$, and the initial position $z_0$, that  reproduce with
the explicit solution the results of the numerical simulation.

\begin{figure}[ht]
\centerline{\epsfig{file=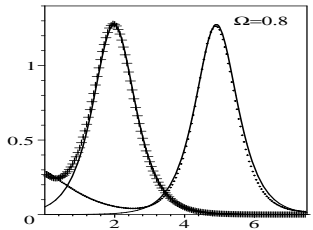,height=3cm,width=5cm}}
\caption{Fit of the numerical solution (crosses) by analytical 
soliton solution (solid line) at  times $65.7$ and $72.7$, 
for an incident frequency $\Omega=0.8$ and amplitude $A=0.53$. }
\label{fig:fit}\end{figure}
An instance of such a fit is displayed in figure \ref{fig:fit} that shows the
analytic soliton $|e(z,t)|^2$ compared with the numerical simulation for two
fixed values of time. This furnishes the following velocities
\begin{equation}\label{tab}
\begin{tabular}{|c|c|c|c|c|c|c|}
\hline
$\Omega$& 0.95  & 0.9 & 0.8 & 0.7  & 0.6 & 0.5  \\
\hline
$ A $    & 0.28 & 0.38 & 0.53 & 0.65 & 0.75 & 0.83   \\
\hline
\hline
 $ v $  & 0.25 & 0.34 & 0.42 & 0.50  & 0.58 & 0.66  \\
\hline 
\end{tabular}\end{equation}
These are {\em slow light pulses} and naturally, small velocities are obtained 
close to the gap edge where the required energy is small.

\paragraph*{Conclusion.}

The property of a Bragg medium in the nonlinear Kerr regime to sustain
nonlinear supratransmission has allowed us to obtain the analytic expression
\eqref{threshold} that fixes the threshold amplitude of an incident cw-beam in
terms of its departure from the Bragg resonance.

This constitutes a practical tool to investigate switching properties of a
Bragg medium and allows at the same time to understand the process at the
origin of sudden transmissivity of the Bragg mirror. It is the nonlinear
supratansmission that results from a nonlinear instability intrinsic to
boundary value problems \cite{instab}.

We expect our result to be usefull for experiments by the ability of the Bragg
medium to become partly transparent by gap soliton generation for an incident
beam of {\em low energy} if its frequency is chosen close to the gap edge.

\paragraph*{Acknowledgments.}
This study was initiated during a stay of one of us (J.L.) at the department
of applied mathematics, Boulder University. It is a pleasure to aknowledge
invitation and enlighting discussions with M.J. Ablowitz.

This work received support from {\em Funda\c c\~ao para \`a Ci\^encia e \`a 
Tecnologia}, grant BPD/5569/2001, and from {\em A\c coes Integradas 
Luso-Francesas}, grant F-4/03.

\end{document}